\documentclass[%
 reprint,
 amsmath,amssymb,
 aps,
 prl,
]{revtex4-2}

\usepackage[utf8]{inputenc}
\usepackage{graphicx}
\usepackage{dcolumn}
\usepackage{bm}
\usepackage{hyperref}

\begin{document}


\title{Asymmetric Roughness of Elastic Interfaces at the Depinning Threshold}

\author{Esko Toivonen}
\email{esko.toivonen@tuni.fi}
\author{Matti Molkkari}%
\author{Esa Räsänen}%
\author{Lasse Laurson}%
\affiliation{%
Computational Physics Laboratory, Tampere University,
P.O. Box 692, FI-33014 Tampere, Finland
}%





\begin{abstract}
Roughness of driven elastic interfaces in random media is 
typically understood to be characterized by a single roughness 
exponent $\zeta$. We show that at the depinning threshold, due 
to symmetry breaking caused by the direction of the driving force, 
elastic interfaces with local, long-range and mean-field elasticity 
exhibit asymmetric roughness. It is manifested 
as a skewed distribution of the local interface heights, and can 
be quantified by using detrended fluctuation analysis to compute 
a spectrum of local, segment-level scaling exponents. The asymmetry 
is observed as approximately linear dependence of the local 
scaling exponents on the difference of the segment height from 
the mean interface height.
\end{abstract}

\maketitle


{\it Introduction.}
Driven elastic interfaces in quenched random media, including, e.g., domain walls in ferromagnets~\cite{zapperi1998dynamics} and ferroelectrics~\cite{paruch2005domain}, contact lines in wetting~\cite{joanny1984model}, and crack fronts in disordered solids~\cite{laurson2013evolution} exhibit universal dynamical response to external driving forces. These features are linked to an underlying depinning phase transition between pinned and moving phases of the interface at a critical external force~\cite{chauve2000creep, nattermann1992dynamics}, originating from the interplay between quenched disorder, elasticity and an external driving force. In addition to dynamical properties such as interface motion taking place in a sequence of avalanches exhibiting scaling~\cite{zapperi1998dynamics,rosso2009avalanche}, a key feature of elastic interfaces at the depinning threshold is their rough morphology~\cite{rosso2002roughness}. The roughness of an elastic interface with a height profile $h(x)$ 
is typically understood to be characterized by a single roughness exponent $\zeta$, e.g., by considering the scaling of the saturated mean squared interface width $W^2(L)\sim L^{2\zeta}$ with the system size $L$, or that of the two-point correlation function $C(x)=\langle (h(x'+x)-h(x'))^2 \rangle \sim x^{2\zeta}$ along the interface, averaged over pinned interface configurations~\cite{duemmer2007depinning}.

\begin{figure}[t!]
    \centering
    \includegraphics[width=\columnwidth]{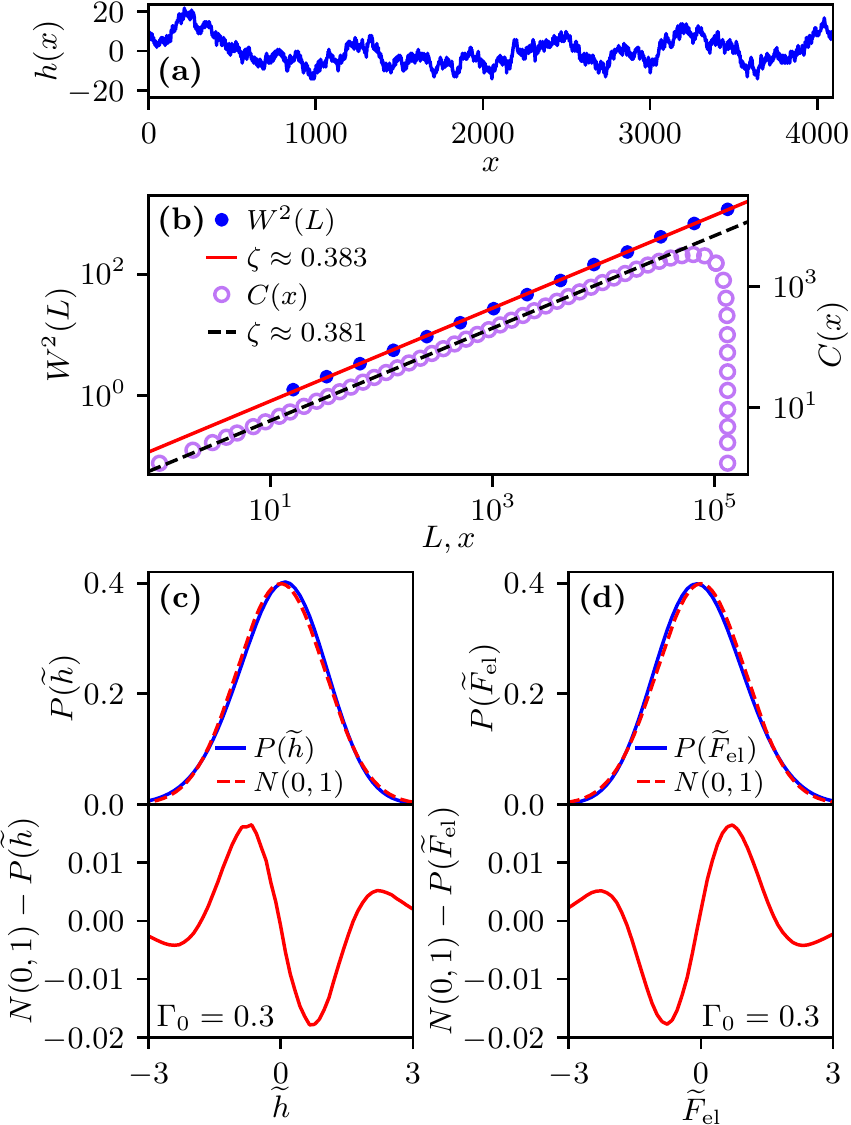}
    \caption{(a) Example of a rough interface configuration $h(x)$ for $L=4096$. (b) Roughness exponent $\zeta$ measured from the scaling of $W^2(L)$ and $C(x)$ (see text). (c) Distribution $P(\tilde{h})$ of the scaled local interface height $\tilde{h} = [h(x)-\langle h \rangle]/\sigma_h$ (top), and the difference between $P(\tilde{h})$ and standard normal distribution $N(0,1)$ (bottom). (d) Distribution $P(\tilde{F}_\mathrm{el})$ of the scaled local elastic force $\tilde{F}_\mathrm{el} = F_\mathrm{el}/\sigma_{F_\mathrm{el}}$ (top), and the difference between $P(\tilde{F}_\mathrm{el})$ and standard normal distribution $N(0,1)$ (bottom).}
    \label{fig:1}
\end{figure}

This simple description assumes that a single roughness exponent sufficiently characterizes the pertinent properties of the system. Thus, this description does not account for any possible asymmetries of $h(x)$ with respect to, e.g., its mean value $\langle h \rangle$.
However, many driven elastic interfaces in quenched random media exhibit local statistical properties and correlations that may greatly diverge from such a simple, symmetric picture. 
Consider as an example a dislocation line (with a rather peculiar non-local self-interaction kernel~\cite{zapperi2001depinning}) driven by an applied shear stress through a sparse set of precipitate particles acting as localized obstacles for dislocation motion~\cite{mohles1999simulation,bako2008dislocation}: the dislocation will bow out in between the precipitates while remaining pinned at them, resulting in noticeable differences between the appearance (e.g., magnitude of the local curvature) of dislocation line segments that have moved more or less than the average dislocation displacement (see, e.g., Fig. 3 of Ref.~\cite{bako2008dislocation}). Hence, by looking at a pinned dislocation configuration at a scale comparable to the precipitate spacing, it is immediately clear which way the dislocation line is being driven by the stress. On the other hand, when observing such interfaces on length scales exceeding the disorder correlation length (as is usually done by construction in simple models of interface depinning), any possible asymmetry with respect to $\langle h \rangle$ is less apparent, see Fig.~\ref{fig:1}(a). 

Here we show that even for scales well above the disorder correlation length where the interface roughness emerges as a consequence of weak or collective pinning~\cite{tanguy1998individual}, 
the roughness of elastic interfaces at the depinning threshold exhibits several asymmetric features, originating from the broken symmetry caused by the direction of the external driving force. Considering as an example system the long-range elastic string~\cite{gao1989first,schmittbuhl1995interfacial,ramanathan1997dynamics} (in what follows we'll use the terms 'string' and 'interface' interchangeably), known to describe systems such as planar cracks~\cite{laurson2013evolution, laurson2010avalanches,bonamy2008crackling}, contact lines~\cite{joanny1984model} and low-angle grain boundaries~\cite{moretti2004depinning}, we find skewed distributions of both the local interface height and the local elastic force; analogous results for local and mean-field elasticity are presented in Supplemental Material~\cite{SM}. 
Analyzing interface segments on different scales conditioned on 
the deviation of the average segment height from the mean interface height and the average elastic force acting on the segment reveals clear trends in the segment height profiles. Hence, we employ detrended fluctuation analysis~\cite{Peng1994dfa,Kantelhardt2001dfa} (DFA) with a scale-dependent segmentation scheme~\cite{Molkkari2020running} to analyze the scaling properties of such segments.
The resulting scaling exponents $\alpha$ are found to converge in the limit of high order of the detrending polynomial, with the converged exponents exhibiting a clear dependence on the difference of the segment height and the mean interface height, and on the average elastic force. The 
average values of these local exponents tend towards the roughness exponent $\zeta$ in the large-scale limit.
Thus, instead of the classical description in terms of a single roughness exponent $\zeta$, our results show that elastic interfaces at the depinning threshold should be characterized by a spectrum of local, segment-level exponents which depend on quantities like the deviation of the segment height from the mean interface height, and/or the average elastic force acting on the segment.

{\it Model: Long-range elastic string.}
We perform extensive simulations of a discretized version of the long-range elastic string in a quenched random medium.  
The local total force acting on the interface element $i$ located at $x=x_i \equiv i$ (with $i$ an integer from 0 to $L$) along the interface $h(x)=h(x_i) \equiv h_i$ is
\begin{equation}
F(x_i) = F_{\mathrm{el}}(x_i) + \eta(x_i,h_i) + F_\mathrm{ext},
\end{equation}
where the first term on the RHS, $F_{\mathrm{el}}(x_i) = \Gamma_0 \sum_{j \neq i} \frac{h_j-h_i}{|x_j-x_i|^2}$ (with $\Gamma_0$ the stiffness of the interface),
represents the long-range elastic interactions, $\eta$ is uncorrelated quenched disorder and $F_\mathrm{ext}$ is the external driving force~\cite{laurson2013evolution}. The parallel dynamics of the interface is defined in discrete time $t$ by setting the local velocity $v(x_i,t) \equiv h(x_i,t+1)-h(x_i,t) = \theta[F(x_i)]$,
where $\theta$ is the Heaviside step function. We employ quasistatic constant velocity driving which keeps the interface in the immediate proximity of the depinning threshold, such that avalanches are triggered by increasing $F_\mathrm{ext}$ just enough to make exactly one interface element unstable [that is, $F(x_i)>0$ for some $i$] whenever the previous avalanche has ended. During an avalanche, $F_\mathrm{ext}$ is decreased at a rate proportional 
to the instantaneous avalanche velocity, $\dot{F}_\mathrm{ext} = -K/L \sum_i v_i (t)$, where $K$ is a parameter controlling the cutoff of the avalanche size distribution~\cite{laurson2013evolution}. To collect data for studying the interface roughness, we simulate the system according to the above protocol, and store interface configurations $h(x)$ from the steady state at regular intervals separated by long enough interface displacements such that consecutive interface configurations are uncorrelated.  
The parameters are set to $L = 4096$, $K = 0.0033$, and $\Gamma_0 = 0.3$ unless stated otherwise, but we consider also different $L$'s up to $L=131072=2^{17}$, and adjust $K$ accordingly to approximately fix the ratio of the correlation length (maximum lateral extent of avalanches) and $L$. 

\begin{figure}[t!]
    \centering
    \includegraphics[width=\columnwidth]{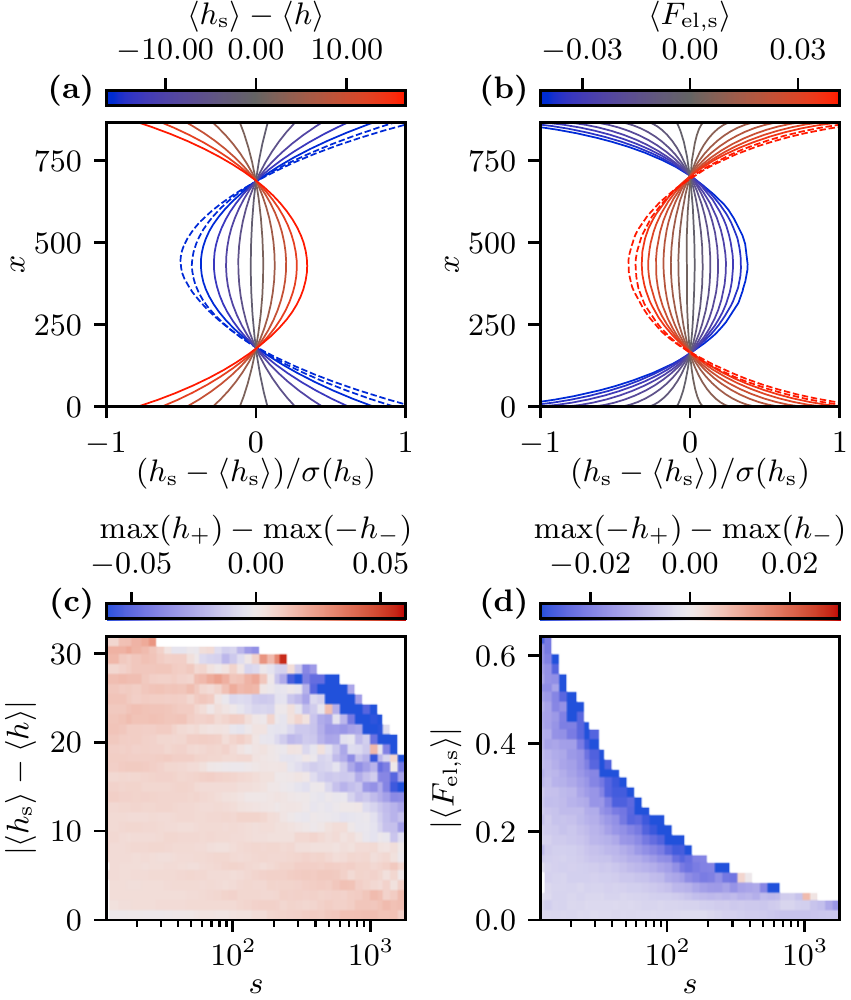}
    \caption{(a-b) Scaled segment profiles $(h_s(x)-\langle h_s \rangle)/\mathrm{std}(h_s)$ for $s=867$ averaged within bins symmetric with respect to $\langle h_s \rangle-\langle h \rangle=0$ and $\langle F_{\mathrm{el},s} \rangle=0$, respectively (solid lines), with the values of $\langle h_s \rangle-\langle h \rangle$ and $\langle F_{\mathrm{el},s} \rangle$ indicated by the colorbars. The dashed lines illustrate the excess profiles for large negative values of $\langle h_s \rangle-\langle h \rangle$ and large positive values of $\langle F_{\mathrm{el},s} \rangle$, respectively. (c-d) Difference between the maxima of the average scaled segment profiles with a positive and the corresponding negative value of $\langle h_s \rangle-\langle h \rangle$ and $\langle F_{\mathrm{el},s} \rangle$, respectively, for different scales $s$.
    }
    \label{fig:2}
\end{figure}

{\it Skewed distribution of interface heights.}
Figure~\ref{fig:1}(a) shows an example of an interface profile $h(x)$ for $L=4096$, illustrating the typical rough morphology one observes above the length scale of the disorder correlation length (which here equals 1). Our interfaces follow the standard scaling picture in that the roughness exponent $\zeta$, estimated from $W^2(L)$ and $C(x)$ in Fig.~\ref{fig:1}(b) for system sizes up to $L=2^{17}$, is very close to the literature value of $\zeta \approx 0.385$~\cite{duemmer2007depinning}. However, a closer look reveals the first signature of asymmetry in the statistical properties of the $h(x)$'s: The distribution of local interface heights, $P(\tilde{h})$, where $\tilde{h} \equiv [h(x)-\langle h \rangle]/\sigma_h$ (with the standard deviation $\sigma_h$ calculated over the whole dataset of $h(x)-\langle h \rangle$ -values), exhibits a small but clearly non-zero negative skewness of -0.183, and hence deviates from the standard normal distribution $N(0,1)$ [Fig.~\ref{fig:1}(c)]. Similar conclusions can be made by considering the distribution of local elastic forces $P(\tilde{F}_\mathrm{el})$ [Fig.~\ref{fig:1}(d), with $\tilde{F}_\mathrm{el} \equiv F_\mathrm{el}(x) / \sigma_{F_\mathrm{el}}$], which exhibits a positive skewness of 0.176. The interpretation of this is that strongly pinned points of the interface lagging behind the mean interface height give rise to a long negative tail in $P(\tilde{h})$, and a long positive tail in $P(\tilde{F}_\mathrm{el})$ as the points $x$ with negative $h(x)-\langle h \rangle$ lagging behind the rest of the interface typically experience a positive $F_\mathrm{el}(x)$. These features are a consequence of the broken symmetry between parts of the interface above and below $\langle h \rangle$ due to the direction of $F_\mathrm{ext}$, and can be reproduced also for interfaces with local and infinite-range (mean-field) interactions (Supplemental Fig.~1), and for continuous-time dynamics (Supplemental Fig.~3~\cite{SM}).

{\it Asymmetric trends in interface segments.}
Next, we examine if the broken symmetry is manifested in other properties of $h(x)$ as well. To this end, we consider interface segments $h_s(x)$ of length $s$,
i.e., we study the problem on various scales $s$, as a function of segment-level quantities such as $\langle h_s \rangle - \langle h \rangle$ (difference of the mean segment height and the mean interface height) and $\langle F_{\mathrm{el},s} \rangle$ (the mean elastic force acting on the segment). Figure~\ref{fig:2}(a) shows a set of ensemble-averaged scaled segment profiles $(h_s(x)-\langle h_s \rangle)/\mathrm{std}(h_s)$ for $s=867$ (a ``large'' example scale smaller than $L$) 
for different values of $\langle h_s \rangle - \langle h \rangle$ [colorbar in Fig.~\ref{fig:2}(a)]. These exhibit clear, approximately parabolic trends for large values of $|\langle h_s \rangle - \langle h \rangle|$, such that the ``opening direction'' of the curves is towards the mean interface height. Moreover, these profiles exhibit asymmetry with respect to $\langle h \rangle$, such that there is an excess of profiles with a large negative $\langle h_s \rangle - \langle h \rangle$, shown as dashed lines in Fig.~\ref{fig:2}(a) in addition to the profiles computed with symmetric binning on both sides of $\langle h_s \rangle - \langle h \rangle = 0$ (solid lines).
Moreover, comparing the symmetrically binned average segment height profiles for different $|\langle h_s \rangle - \langle h \rangle|$ and $s$, by computing the difference $\mathrm{max}(h_+)-\mathrm{max}(-h_-)$ [where $\mathrm{max}(h_+)$ refers to the maximum of the normalized average segment height profile with a positive $\langle h_s \rangle - \langle h \rangle$, and $\mathrm{max}(-h_-)$ is the maximum of the negative normalized average segment height profile with a negative $\langle h_s \rangle - \langle h \rangle$; only bins with more than 10000 segments are considered here to avoid spurious effects due to statistical noise] reveals an additional signature of asymmetry: For small $s$, the difference is close to zero but slightly positive [pale red in Fig.~\ref{fig:2}(c)], while for large $s$ and $|\langle h_s \rangle - \langle h \rangle|$ it becomes clearly negative [blue in Fig.~\ref{fig:2}(c)], showing that the interface segments exhibit asymmetry also for large scales, in addition to the long negative tail in the distribution $P(\tilde{h})$ of local heights in Fig.~\ref{fig:1}(c). 

Analogous quantities can be studied by considering segments with different average elastic forces $\langle F_{\mathrm{el},s} \rangle$ [Figs.~\ref{fig:2}(b) and (d)]. Figure~\ref{fig:2}(b) shows a set of ensemble-averaged segment profiles for $s=867$ for different values of $\langle F_{\mathrm{el},s} \rangle$ [colorbar in Fig.~\ref{fig:2}(b)]. These average segment profiles are qualitatively similar to the ones found above when conditioning with the value of $\langle h_s \rangle - \langle h \rangle$, but the trends are somewhat less parabolic-looking, suggesting that they may be better captured by a higher-order polynomial. Again, there is an excess of profiles with a large positive value of $\langle F_{\mathrm{el},s} \rangle$ (dashed lines in Fig.~\ref{fig:2}(b)), corresponding to segments which are lagging behind the mean height of the interface. Another difference is that for large $s$, $\langle F_{\mathrm{el},s} \rangle$ has a tendency to approach zero, and hence the interval of $|\langle F_{\mathrm{el},s}\rangle|$-values in Fig.~\ref{fig:2}(d) gets increasingly narrow as larger $s$'s are considered.
Nevertheless, an analogous asymmetry is seen also in Fig.~\ref{fig:2}(d), such that for small scales $\mathrm{max}(-h_+)-\mathrm{max}(h_-)$ is close to zero ($+$ and $-$ now refer to positive and negative values of $\langle F_{\mathrm{el},s} \rangle$, respectively), while a clearly negative value is found for the largest $s$'s for a given $|\langle F_{\mathrm{el},s}\rangle|$. These findings constitute a large-scale analog of the long positive tail in the distribution $P(\tilde{F}_\mathrm{el})$ of local elastic forces in Fig.~\ref{fig:1}(d).

\begin{figure}[t!]
    \centering
    \includegraphics[width=\columnwidth]{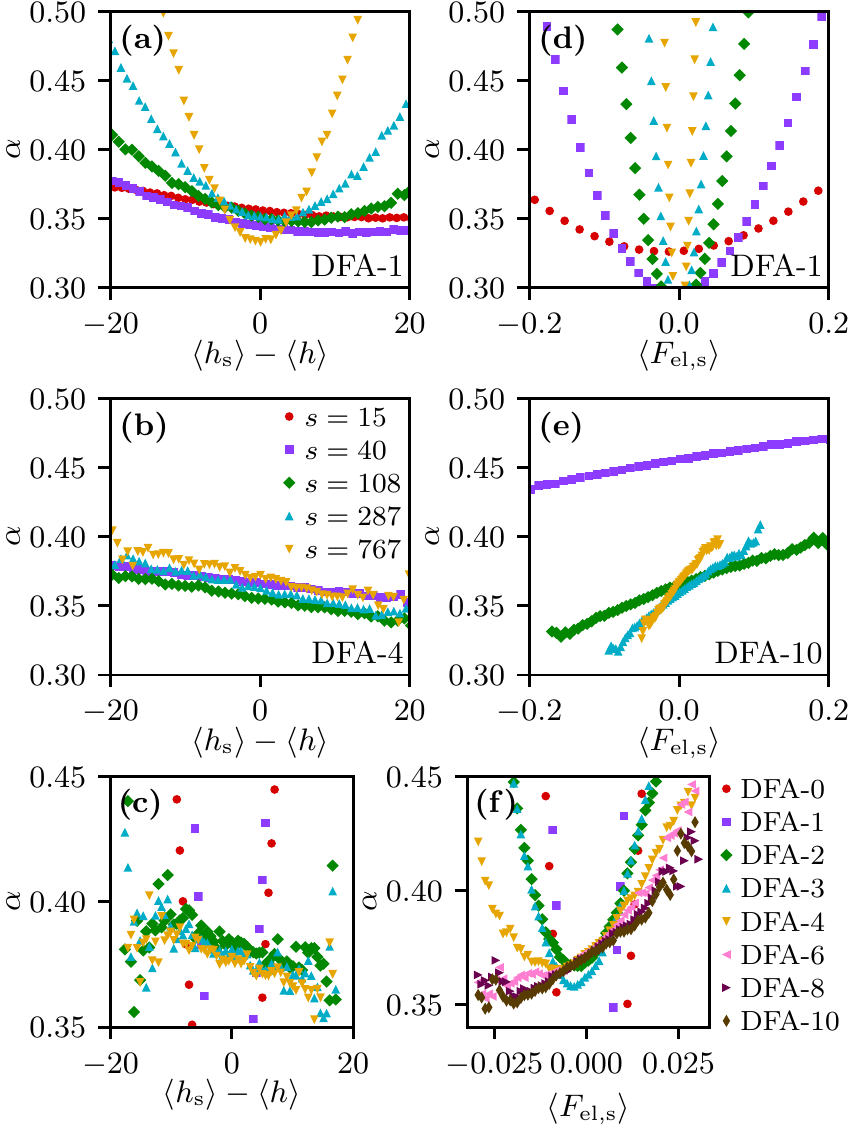}
    \caption{(a-b) DFA scaling exponents $\alpha$ for different $s$ [legend in (b)] as a function of $\langle h_s \rangle - \langle h \rangle$ for DFA-1 and DFA-4, respectively. (c) Convergence of the $\alpha$'s for different DFA orders for $s=1602$. (d-f) Corresponding data as a function of $\langle F_{\mathrm{el},s} \rangle$, with DFA-10 in (e).
    }
    \label{fig:3}
\end{figure}

{\it Scaling properties of the segments.}
Next, we address the question of the scaling properties of the segments $h_s(x)$, and how these may depend on $\langle h_s \rangle - \langle h \rangle$ and $\langle F_{\mathrm{el},s} \rangle$. Given the clear trends illustrated in Figs.~\ref{fig:2}(a) and (b), we use DFA-$n$, which performs local detrending with least-squares fitting of $n$-th degree polynomials in windows of length $s$. The mean squared differences from the trends are computed and averaged over all the windows to obtain the squared fluctuation function $F(s)^2$. Conventionally, the scaling exponent $\alpha$ is obtained by a linear fit from a logarithmic plot of $F(s) \propto s^{\alpha}$~\cite{Peng1994dfa,Kantelhardt2001dfa}. Therefore, with DFA-0 the exponents $\alpha$ and $\zeta$ are equal \footnote{The least-squares fit of a 0-th order polynomial is equal to the mean, and hence the mean squared differences become variances. We also omit the integration step usually performed in DFA, as the height profile is already considered a random walk and not its steps.}.
We perform a scale-dependent segmentation scheme~\cite{Molkkari2020running} to obtain scale-dependent exponents $\alpha(s)$ in short segments of the height profiles: The fluctuation function is computed in maximally overlapping windows at scales $s$, $s \pm 1$ and $\alpha(s)$ is obtained by central finite differences from the logarithmic quantities~\cite{Molkkari2020running}.
To achieve maximal spatial locality for our results, we compute the fluctuation functions in segments of length $s+1$. 

Figures~\ref{fig:3}(a) and (d) show the scaling exponent $\alpha$ obtained from DFA-1 for various scales $s$ as functions of $\langle h_s \rangle - \langle h \rangle$ and $\langle F_{\mathrm{el},s} \rangle$, respectively. These exhibit parabolic-like dependencies on $\langle h_s \rangle - \langle h \rangle$ and $\langle F_{\mathrm{el},s} \rangle$, which however are a consequence of the linear detrending not being sufficient here given the higher-order trends revealed in Figs.~\ref{fig:2}(a) and (b). In Figs.~\ref{fig:3}(b) and (e), the corresponding data is shown as obtained using higher order polynomials for detrending [DFA-4 and DFA-10 in Figs.~\ref{fig:3}(b) and (e), respectively, chosen to represent the converged results]. To illustrate the convergence of the results upon increasing the DFA order, Figs.~\ref{fig:3}(c) and (f) show the $\alpha$-values for a fixed $s=1602$ (a "large" example scale), 
obtained by using different orders of the detrending polynomial. 
In the limit of high DFA order we find a key result of this paper, i.e., an approximately linear dependence of $\alpha$ on $\langle h_s \rangle - \langle h \rangle$ and $\langle F_{\mathrm{el},s} \rangle$, with the slope being negative in Figs.~\ref{fig:3}(b) and (c) [height difference], and positive in Figs.~\ref{fig:3}(e) and (f) [elastic force], showing how the broken symmetry due to the external force is manifested in the scaling properties of the interface segments.
Notice how a higher order polynomial is needed for detrending of the segments conditioned on the value of $\langle F_{\mathrm{el},s} \rangle$, consistent with the non-parabolic profiles in Fig.~\ref{fig:2}(b).
We also note that for sufficiently large $s$ (approximately for $s \gtrsim 100$), $\alpha$ does not exhibit any clear dependence on $s$, consistent with the scale-free nature of fluctuations of $h(x)$ expected at the depinning threshold. This is directly evident in Fig.~\ref{fig:3}(b) where the curves for different $s$ approximately overlap. A similar conclusion can be reached regarding the data shown in Fig.~\ref{fig:3}(e) if one rescales the horizontal axis with the $s$-dependent range of $\langle F_{\mathrm{el},s} \rangle$ [not shown, see also Fig.~\ref{fig:2}(d)].
Thus, we generally find a larger $\alpha$ for negative $\langle h_s \rangle - \langle h \rangle$ and positive $\langle F_{\mathrm{el},s} \rangle$, i.e., for segments that are lagging behind the average interface. This is likely due to the ``stretched'' nature of the strongly pinned segments which are being pulled forward by the combination of $F_\mathrm{ext}$ and $\langle F_{\mathrm{el},s} \rangle$ [see also the dashed lines  in Figs.~\ref{fig:2} (a) and (b)], resulting in stronger correlations (larger $\alpha$) in such segments. Analogous results are obtained also for interfaces with local elasticity (Supplemental Fig.~2) and for continuous-time dynamics (Supplemental Fig.~4~\cite{SM}). 

\begin{figure}[t!]
    \centering
    \includegraphics[width=\columnwidth]{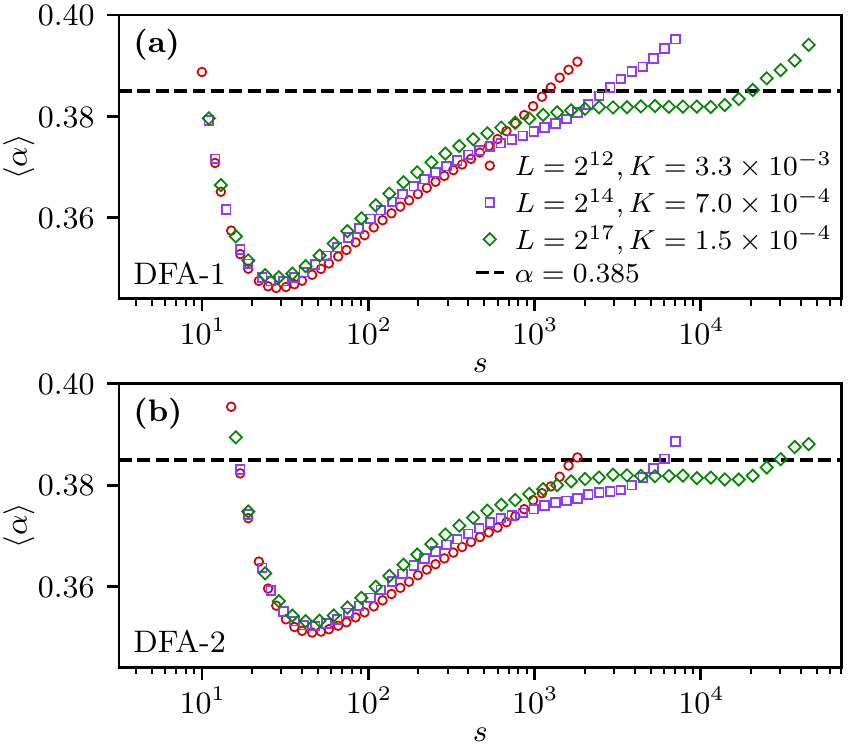}
    \caption{Average scaling exponent $\langle \alpha \rangle$, weighted by the number of occurrences of segments with the different $\langle h_s \rangle - \langle h \rangle$ -values, for different $L$ (and $K$) as a function of the scale $s$, considering DFA-1 and DFA-2 in (a) and (b), respectively. Dashed lines indicate the literature value of the roughness exponent $\zeta \approx 0.385$.}
    \label{fig:4}
\end{figure}

Finally, we consider the relation of these local, segment-level exponents $\alpha$ and the global 
roughness exponent $\zeta$. To this end, in Fig.~\ref{fig:4}, we consider the average of the $\langle h_s \rangle - \langle h \rangle$-dependent DFA scaling exponent $\langle \alpha \rangle$, weighted by the number of occurrences of the different $\langle h_s \rangle - \langle h \rangle$-values in the dataset, which equals the average of $\alpha$ of segments of a given $s$. 
Figures~\ref{fig:4}(a) and (b) show the resulting $\langle \alpha \rangle$'s as a function of the scale $s$ for different system sizes (with the $K$-parameter adjusted so that the correlation length along the interface remains roughly the same fraction of $L$ for different $L$), considering DFA-1 and DFA-2, respectively. In the limit of large $s$, $\langle \alpha \rangle$ converges to a value very close to $\zeta \approx 0.385$ (dashed lines in Fig.~\ref{fig:4}) largely independently of the DFA order (the increase of $\langle \alpha \rangle$ for the very largest $s$'s is likely due to the $K$-dependent correlation length).
Thus, the global roughness exponent emerges on large scales as an average of the local, $\langle h_s \rangle - \langle h \rangle$-dependent scaling exponents.

{\it Conclusions.}
To conclude, our results show that  
$F_\textrm{ext}$ breaks the symmetry of roughness with respect to 
$\langle h \rangle$ of elastic interfaces in random media at the depinning threshold, 
suggesting that a single roughness exponent $\zeta$ is not a full description of their rough morphology, 
and that the spectrum of local, segment-level exponents needs to be considered as well. 
We emphasize that this result applies for all ranges of the elastic interactions, and is true already on the level of {\it individual} interface configurations, and hence this result significantly adds to previous studies arguing that distributions $P(w^2)$ of the interface width are needed in addition to $\zeta$ to characterize {\it ensembles} of rough interface configurations~\cite{rosso2003universal}.
Our results might be relevant for related problems like the scaling properties of anisotropic fracture surfaces~\cite{ponson2006two}, and call for experimental studies 
of diverse systems ranging from domain walls in ferromagnetic thin films~\cite{albornoz2021domain} to planar crack fronts~\cite{santucci2010fracture}. Finally, 
an interesting avenue for future work would be to check if the asymmetry persists in the thermally activated creep regime governed by the equilibrium roughness exponent~\cite{ferrero2021creep}.

\providecommand{\noopsort}[1]{}\providecommand{\singleletter}[1]{#1}%

\end{document}


\title{Supplemental Material for \\ ``Asymmetric Roughness of Elastic
Interfaces at the Depinning Threshold''}
\author{Esko Toivonen}
\author{Matti Molkkari}%
\author{Esa Räsänen}%
\author{Lasse Laurson}%
\maketitle

Here we briefly complement the results of the main article where we focus on the long-range elastic string by presenting results also for strings/interfaces with local and mean-field elasticity, and consider also the effect of varying the elastic stiffness $\Gamma_0$ of the local and non-local models, as well as the effect of the microscopic dynamics (discrete vs continuous) on the asymmetric roughness of the interfaces. 

{\it Models with local and mean-field elasticity.} A description of an elastic string with local elasticity is given by the discretized quenched Edwards-Wilkinson (qEW) equation, obtained by replacing the elastic force term in Eq.~(1) of the main text by
\begin{equation}
F_\textrm{el,qEW}(x_i) = \Gamma_0 \nabla^2 h(x_i) = \Gamma_0 (h_{i+1}+h_{i-1}-2h_i).
\end{equation}
The mean-field limit, on the other hand, is described by infinite-range elastic interactions, 
\begin{equation}
\label{eq:mf}
F_\textrm{el,MF}(x_i) = \Gamma_0 (\langle h \rangle - h_i),    
\end{equation}
where all the line segments $h_i$ interact with the average interface height $\langle h \rangle$. It is worth noting that Eq.~(\ref{eq:mf}) implies that the mean-field model as defined here does not have any spatial structure (since all the line segments interact with the mean interface height $\langle h \rangle$, any spatial order of the $h_i$'s gives rise to the same results), and hence in what follows we consider only the height distribution $P(\tilde{h})$ and its skewness for the mean-field model. Notice also that for the mean-field model $P(\tilde{h})$ and $P(\tilde{F}_\mathrm{el})$ are trivially equivalent apart from the signs of any skewness they might exhibit being opposite, and hence we focus on $P(\tilde{h})$ only. For the qEW equation, we consider also the scaling properties of the rough interfaces similarly to what is done in the context the long-range elastic string in the main text.

\begin{figure}[t!]
	\centering
	\includegraphics[width=\columnwidth]{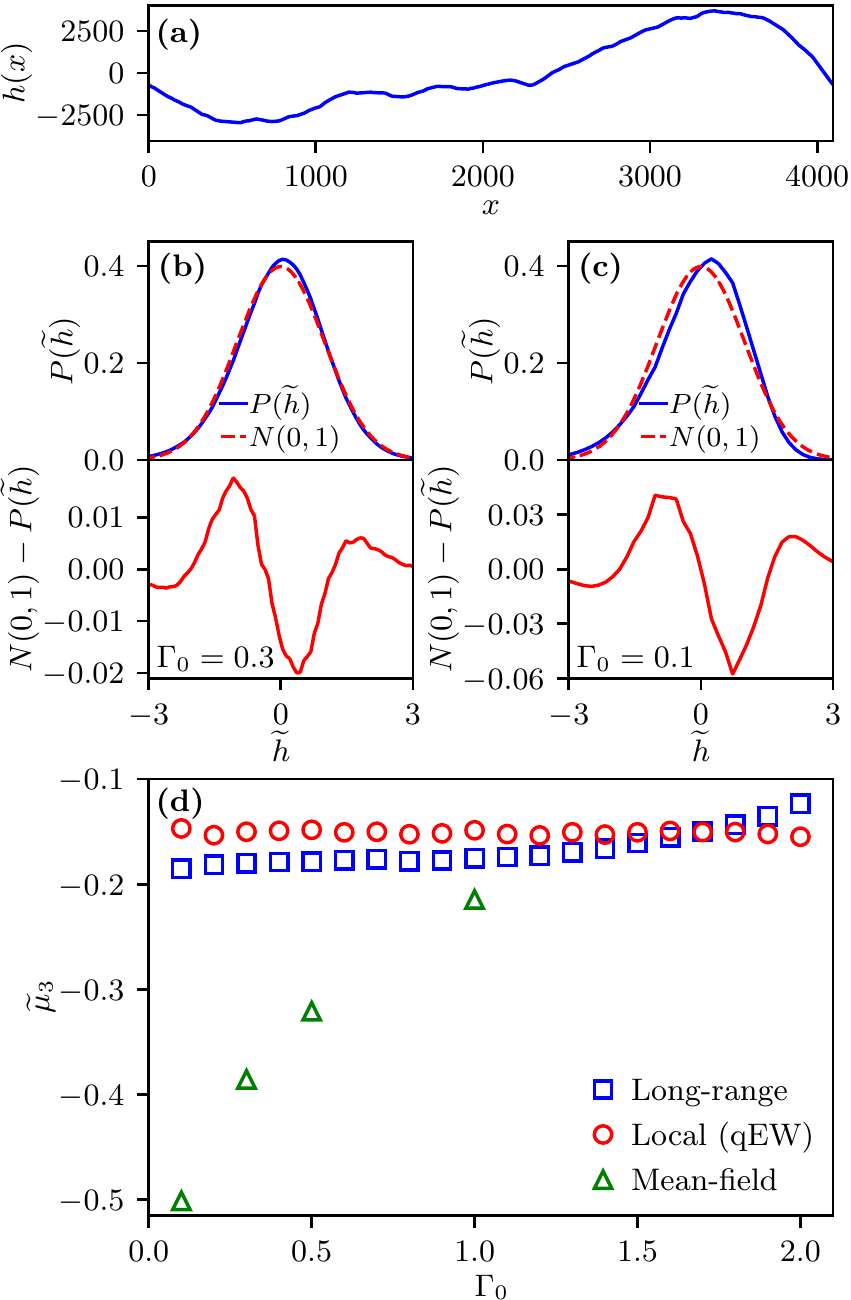}
	\caption{(a) Example of a rough interface configuration $h(x)$ for the qEW equation with $L = 4096$. (b) Distribution $P(\tilde{h})$ of the scaled local interface height $\tilde{h} = [h(x)-\langle h \rangle]/\sigma_h$ (top), and the difference between $P(\tilde{h})$ and standard normal distribution $N(0,1)$ (bottom) for the qEW equation. (c) Same as in (b) but for the mean field model. (d) Dependence of the skewness $\tilde{\mu}_3$ on $\Gamma_0$ for the three models considered.}
	\label{smfig1}
\end{figure}

{\it Skewness of the height distributions.} Supplemental Fig.~\ref{smfig1}(a) shows an example line profile $h(x)$ from simulations of the qEW equation with $L=4096$; notice the significant difference in the overall appearance of the profile as compared to that shown in Fig.~1(a) of the main text for the long-range elastic string. Supplemental Fig.~\ref{smfig1}(b) shows the distribution of the scaled local interface heights $P(\tilde{h})$ for $\Gamma_0=0.3$; comparison with the standard normal distribution reveals again noticeable negative skewness $\tilde{\mu}_3$, which is shown as a function of $\Gamma_0$ in Supplemental Fig.~\ref{smfig1}(d) (red circles). For the qEW equation (local elasticity) $\tilde{\mu}_3$ fluctuates close to $\tilde{\mu}_3 \approx -0.15$ largely independent of $\Gamma_0$. The corresponding data for the long-range elastic string (discussed in the main text) exhibits some dependence on $\Gamma_0$ such that smaller $\Gamma_0$ (corresponding to stronger pinning in relative terms) leads to more negative skewness. Nevertheless, $\tilde{\mu}_3$ is clearly negative for the entire range of $\Gamma_0$-values considered.

 Supplemental Fig.~\ref{smfig1}(c) shows the distribution of the scaled local interface heights $P(\tilde{h})$ in the case of mean-field (infinite range) elasticity with $\Gamma_0=0.1$, revealing again noticeable negative skewness. The skewness $\tilde{\mu}_3$ of the mean field system exhibits a rather strong dependence on $\Gamma_0$ [green triangles in Supplemental Fig.~\ref{smfig1}(d)], such that smaller $\Gamma_0$ (stronger pinning) results in more pronounced negative skewness. One may note that considering a fixed $\Gamma_0<1$ for the different models with different range of elastic interactions, the skewness increases in magnitude with increasing interaction range. Overall, all the three systems/universality classes considered consistently exhibit negative skewness of the the distribution of the scaled local interface heights $P(\tilde{h})$.

\begin{figure}[t!]
	\centering
	\includegraphics[width=\columnwidth]{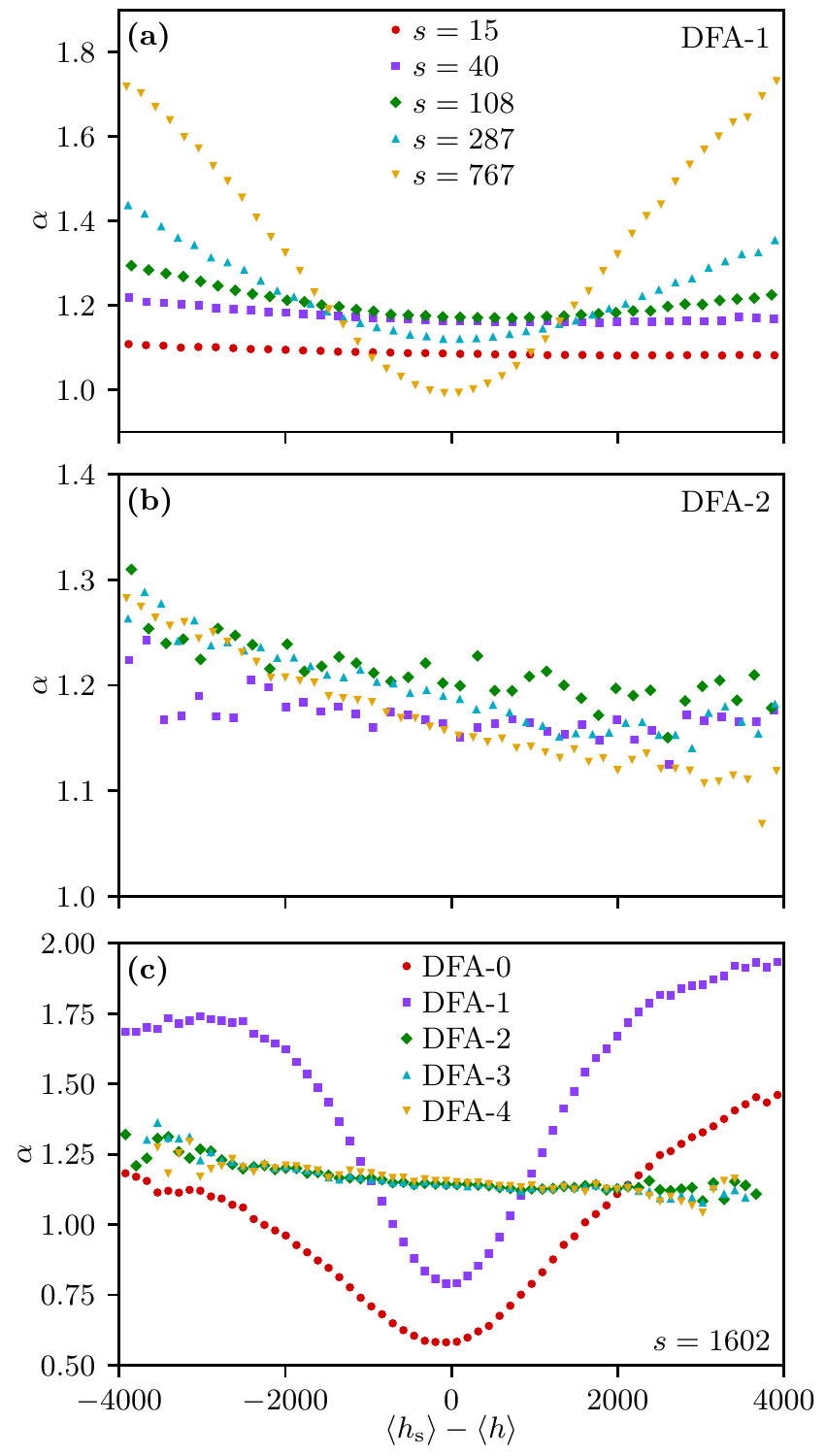}
	\caption{Scaling properties of the segments for the qEW equation (local elasticity). (a-b) DFA scaling exponents $\alpha$ for different $s$ [legend in (a)] as a function of $\langle h_s \rangle - \langle h \rangle$ for DFA-1 and DFA-2, respectively. (c) Convergence of the $\alpha$’s for different DFA orders for $s = 1602$. }
	\label{smfig2}
\end{figure}

\begin{figure}[t!]
	\centering
	\includegraphics[width=\columnwidth]{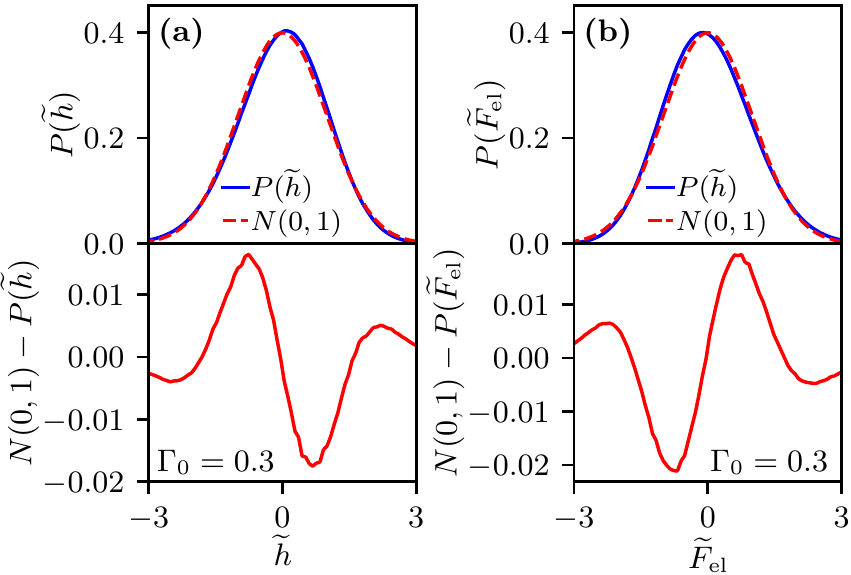}
	\caption{Distribution $P(\tilde{h})$ of the scaled local interface height $\tilde{h} = [h(x)-\langle h \rangle]/\sigma_h$ (top, skewness~$-0.17$), and the difference between $P(\tilde{h})$ and standard normal distribution $N(0,1)$ (bottom) for the continuous-time long-range elastic string with $L = 1024$ (b) Distribution $P(\tilde{F}_\mathrm{el})$ of the scaled local elastic force $\tilde{F}_\mathrm{el} = F_\mathrm{el}/\sigma_{F_\mathrm{el}}$ (top, skewness $0.21$), and the difference between $P(\tilde{F}_\mathrm{el})$ and standard normal distribution $N(0,1)$ (bottom).}
	\label{smfig3}
\end{figure}

{\it Scaling properties of the interface segments for local elasticity.} Supplemental Fig.~\ref{smfig2} summarizes the scaling properties of the rough lines in the case of the qEW equation/local elasticity; the data shown is analogous to Figs.~3(a), (b) and (c) of the main text where the long-range elastic string is considered. Supplemental Fig.~\ref{smfig2}(a) shows the scaling exponent $\alpha$ obtained from DFA-1 for various scales $s$ as a function of $\langle h_s\rangle - \langle h \rangle$. Similarly to Fig.~3(a) of the main text, these exhibit parabolic-like dependencies on $\langle h_s\rangle - \langle h \rangle$, which we again attribute to linear detrending not being able to remove the trends present in the data. Supplemental Fig.~\ref{smfig2}(b) shows the corresponding data from DFA-2, chosen to represent the approximately converged results for a wide range of scales $s$. 
This convergence with increasing DFA order is further illustrated in Supplemental Fig.~\ref{smfig2}(c) for a large example scale of $s=1602$, where $\alpha$-values obtained from DFA-2 and above fall on top of the same linear trend line close to the known value of the roughness exponent $\zeta_\mathrm{qEW} \approx 1.25$~\cite{ferrero2013nonsteady,rosso2003depinning}. The linear dependence of $\alpha$ on $\langle h_s\rangle - \langle h \rangle$ is again a manifestation of asymmetric roughness. 

\begin{figure}[t!]
	\centering
	\includegraphics[width=\columnwidth]{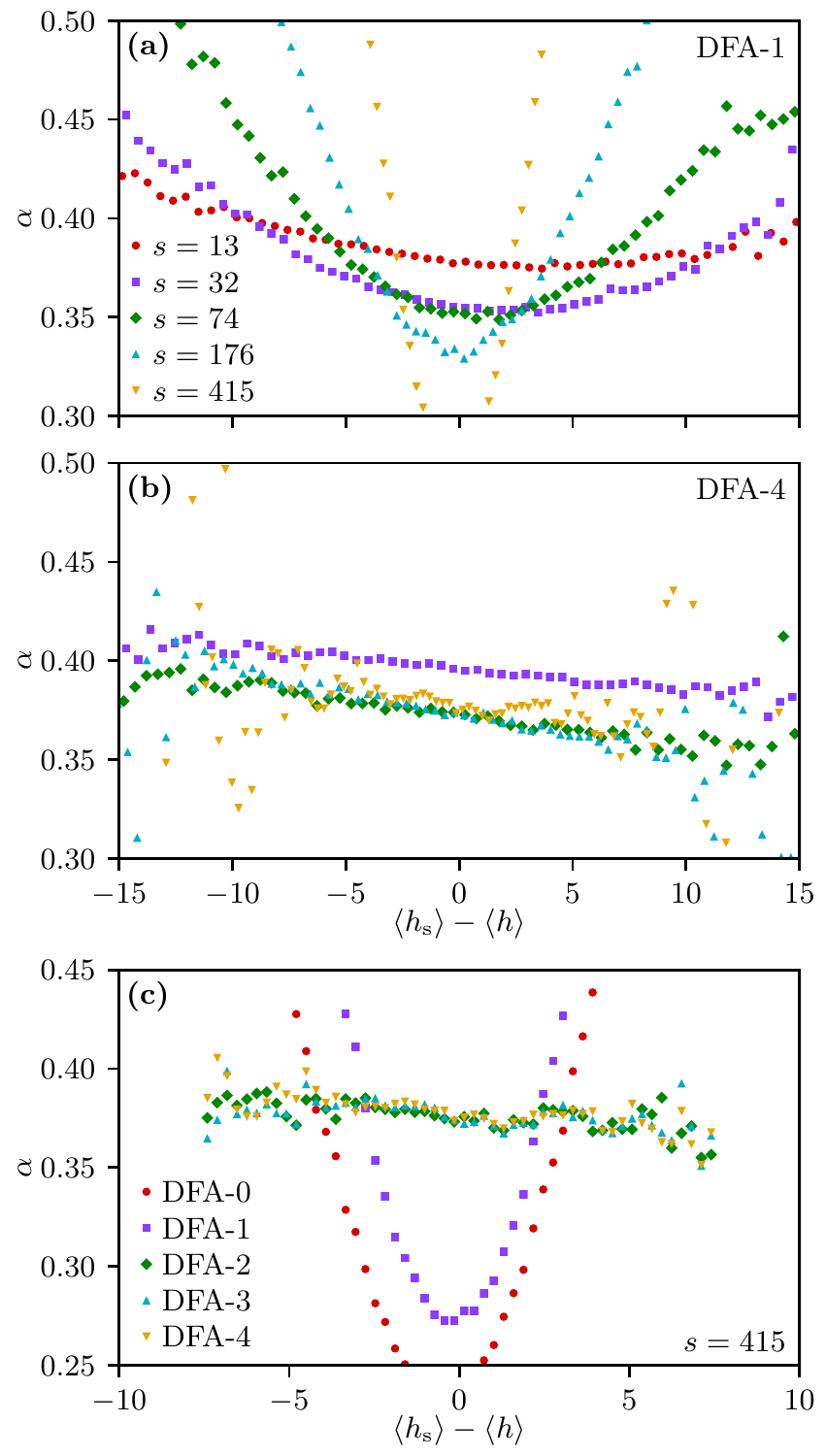}
	\caption{Scaling properties of interface segments for the continuous-time long-range elastic string for different scales $s$ as obtained from DFA-1 (a) and DFA-4 (b), and for different orders of DFA for a fixed $s=415$.}
	\label{smfig4}
\end{figure}

{\it Long-range elastic string with continuous-time dynamics.} Finally, we address the possible role of the microscopic dynamics on the observed asymmetry. In the time-discretized models considered above and in the main text, the local interface velocity at the time step $t$ is given by $v(x_i,t) \equiv \theta[F(x_i)]$, where  $\theta$ is the Heaviside step function. The corresponding continuous-time models are defined via a linear mobility law, where the local instantaneous interface velocity is proportional to the total force acting on the interface segment, i.e., the interface velocity at the time step $t$ at $x_i$ is given by $v(x_i,t) \propto F(x_i)$, with a continuous disorder field $\eta(x_i,h)$ defined for each $x_i$ by employing spline interpolation. Supplemental Fig.~\ref{smfig3} shows the resulting distributions of scaled interface heights $P(\tilde{h})$ and elastic forces $P(\tilde{F}_\mathrm{el})$ for the continuous-time long-range elastic string. These are practically identical to those shown in Fig.~1 of the main text for the corresponding discrete model, with the skewness values of -0.17 and 0.21 for $P(\tilde{h})$ and $P(\tilde{F}_\mathrm{el})$, respectively. Supplemental Fig.~\ref{smfig4} shows the data from the continuous time model corresponding to that shown in Fig.~3 of the main text for the discrete model. Again, the results are essentially the same for both the discrete and continuous model, such that the asymmetric scaling properties are present also in the continuous-time model. Thus, the asymmetry we report is not a consequence of the chosen microscopic dynamics, but is a general property of driven interfaces at the depinning threshold.

To sum up, the broken symmetry reported in the main text for the physically interesting case of the long-range elastic string is present also in strings governed by local and mean field elasticity, as well as in models with discrete and continuous time dynamics, thus providing strong evidence for the general nature of the asymmetric roughness of elastic interfaces at the depinning threshold.

\bibliographystyle{apsrev4-2}
%